\documentclass[aps,preprint,amsmath,amssymb,superscriptaddress]{revtex4}
\usepackage{amsmath,amssymb}
\usepackage{graphicx}
\usepackage{booktabs}
\usepackage[usenames, dvipsnames]{color}
\usepackage{multirow}
\usepackage{setspace}
\usepackage{rotating}
\usepackage[ pdftex, plainpages = false, pdfpagelabels, 
                 pdfpagelayout = useoutlines,
                 bookmarks,
                 bookmarksopen = true,
                 bookmarksnumbered = true,
                 breaklinks = true,
                 linktocpage=all,
                 pagebackref=false,
                 colorlinks = true,
                 linkcolor = BrickRed,
                 urlcolor  = blue,
                 citecolor = BrickRed,
                 anchorcolor = green,
                 hyperindex = true,
                 hyperfigures
                 ]{hyperref} 

\begin{document}
\bibliographystyle{utphys}

\title{\href{http://necsi.edu/research/evoeco/spatialevolution.html}{Eco-Evolutionary Feedback in Host--Pathogen Spatial Dynamics}}
\author{\href{http://www.sunclipse.org}{Blake C.\ Stacey}}
\affiliation{Martin A.\ Fisher School of Physics, Brandeis University}
\affiliation{New England Complex Systems Institute}
\author{Andreas Gros}
\affiliation{New England Complex Systems Institute}
\author{\href{http://www.necsi.edu/faculty/bar-yam.html}{Yaneer Bar-\!Yam}}
\affiliation{New England Complex Systems Institute}

\date{\today}

\begin{abstract}
Spatial extent is a complicating factor in mathematical biology.  The
possibility that an action at point A cannot immediately affect what
happens at point B creates the opportunity for spatial nonuniformity.
This nonuniformity must change our understanding of evolutionary
dynamics, as the same organism in different places can have different
expected evolutionary outcomes. Since organism origins and fates are
both determined locally, we must consider heterogeneity explicitly to
determine its effects. We use simulations of spatially extended
host--pathogen and predator--prey ecosystems to reveal the limitations
of standard mathematical treatments of spatial heterogeneity. Our
model ecosystem generates heterogeneity dynamically; an adaptive
network of hosts on which pathogens are transmitted arises as an
emergent phenomenon. The structure and dynamics of this network differ
in significant ways from those of related models studied in the
adaptive-network field. We use a new technique, organism swapping, to
test the efficacy of both simple approximations and more elaborate
moment-closure methods, and a new measure to reveal the timescale
dependence of invasive-strain behavior.  Our results demonstrate the
failure not only of the most straightforward (``mean field'')
approximation, which smooths over heterogeneity entirely, but also of
the standard correction (``pair approximation'') to the mean field
treatment. In spatial contexts, invasive pathogen varieties can
prosper initially but perish in the medium term, implying that the
concepts of reproductive fitness and the Evolutionary Stable Strategy
have to be modified for such systems.
\end{abstract}

\maketitle

\section{Introduction}
\label{sec:intro}
Mathematical modeling of biological systems involves a tradeoff
between detail and tractability.  Here, we consider evolutionary
ecological systems with spatial extent---a complicating factor.
Analytical treatments of spatial systems typically treat as equivalent
all configurations with the same overall population density, the same
allele frequencies, the same pairwise contact probabilities or the
like.  For ease of analysis, one seeks a simplified analytical model,
which {\em coarse-grains} ``microstates'' (the complete specification
of each organism) to ``macrostates'' (characterized by quantities like
average densities), allowing one to make useful predictions about the
model's behavior~\cite{levin1992,dieckmann2000}.  Corrections to
simple coarse-grainings can quickly generate an overbearing quantity
of algebra.  It is fairly well appreciated that the simplest
approximations break down in the spatial context.  What is less
acknowledged and not yet systematically understood is that the
extensions of the simpler approximations also fail.  Before exhausting
ourselves with ever-more-elaborate refinements, it would be useful to
have some understanding of when a particular series of approximations
is doomed to inadequacy.

In this article, we study the context in which commonly-used
coarse-grainings can be expected to fail at capturing the evolutionary
dynamics of an ecosystem, and in addition we provide a novel, direct
demonstration of that failure.  The fundamental issue is {\em spatial
  heterogeneity,} a long-recognized concern for mathematical
biology~\cite{wright1945, hartl2007}.  When does spatial heterogeneity
significantly impact the choice of appropriate mathematical treatment,
and when does a chosen mathematical formalism not capture the full
implications of spatial variability?  We show that one can test a
treatment of heterogeneity by transplanting organisms within a
simulated ecosystem in such a way that, were the treatment valid, the
modeled behavior of the ecosystem over time would remain essentially
unchanged.  We demonstrate situations where the system's behavior
changes dramatically and cannot be captured by a conventional
treatment.  The complications we explore imply that {\em short-term
  descriptions} of what is happening in an evolutionary ecological
model can be insufficient and, in fact, misleading, with regard not
just to quantitative details but also to qualitative characteristics
of ecological dynamics.

Many modeling approaches in mathematical biology which appear distinct
at first glance turn out to be describing the same phenomenon with
different equations~\cite{page2002, bijma2008, damore2011}.  What matters
for our purposes is not so much which technique is chosen, but whether
the underlying assumptions do, in fact, apply.

``Mean-field theory'' is a term from statistical
physics~\cite{baryam1999, kardar2007} which has been adopted in
ecology~\cite{goodnight2008, lion2008, givan2011}, referring to an
approximation in which each component of a system is modeled as
experiencing the same environment as any other.  This implies that the
probability distribution over all possible states of the system
factors into a product of probability distributions for individual
components.  An example in population genetics is the assumption that
a population is panmictic.  That is, if a new individual in one
generation has an equal chance of receiving an allele from any
individual in the previous generation, then we can approximate the
ecosystem dynamics using only the proportion of that allele, rather
than some more complicated representation of the population's genetic
makeup.  Modeling evolution of that population as ``change in allele
frequencies over time'' (per, {\em e.g.,}~\cite{williams1992,
  page2002}) is, implicitly, a mean-field
approximation~\cite{sayama2000}.  The mean-field approximation is also
in force if one postulates that an individual organism interacts with
some subset, chosen at random, of the total population, even if the
form and effect of interactions within that subset are complicated (as
in, {\em e.g.,}~\cite{vandyken2011, archetti2011}).

It is well known that real species are not necessarily panmictic.
However, many treatments which acknowledge this are still mean-field
models.  The textbook way of incorporating geographical distance into
a population-genetic model is to divide the system into $N$ local
subpopulations, ``islands,'' connected via migration~\cite{taylor1996,
  kokko2008, wild2009}.  Within each subpopulation, distance is
treated as negligible, and organisms are well mixed~\cite{levins1969,
  hartl2007}.  This approach makes a simplifying assumption that there
is a single distance scale below which panmixia
prevails~\cite{platt2010}, and it relies on well-defined boundaries
between panmictic subpopulations which persist over
time~\cite{levins1969}.  Furthermore, the connections among
subpopulations are frequently taken to have the topology of a complete
graph, {\em i.e.,} an organism in one subpopulation can migrate to any
other with equal ease~\cite{levins1969, hartl2007, kokko2008,
  wild2009}.  In this case, each of the $N$ subpopulations do
experience the same environment, to within one part in $N$.  Thus, the
mean-field approximation is in force at the island level, and the
island model incorporates spatial extent without incorporating a full
treatment of spatial heterogeneity.  For real
ecosystems~\cite{halley2004, scanlon2007, platt2010, reigada2012}, one
or more of these simplifying assumptions can fail.  Long-distance
migration is often thought to return a spatial ecosystem to a
well-mixed form, but if organisms' migration habits are themselves
adaptive, this is not necessarily so~\cite{ichinose2013}.  More
complicated population structures require more sophisticated
mathematical treatments of evolution, a fact which has mathematical
consequences, but more importantly has real-world implications for
practical issues like the evolution of drug-resistant
diseases~\cite{escalante2009}.

Where mean field approximations fail, ``higher order'' approximations
may be employed.  Rather than individual organisms or islands, a pair
approximation considers pairs of organisms or pairs of spatial regions
in average contexts.  However, this approximation can also fail when
local contexts of groups do not reflect the overall system behavior
due to heterogeneity across larger domains.  Patches of distinct
genetic composition in different parts of a spatial system that are
well separated cannot be treated correctly by such approximations.
Quantitative analyses confirm this inadequacy.  We introduce a new
approach to analyzing such approximations by swapping pairs of
organisms in a way that preserves the pair description. For spatial
systems, such swapping events violate the spatial separation between
patches and changes the evolutionary behavior of the system. The
swapping method therefore serves as a direct test of the (in)adequacy
of the pair approximation.  For evolution on random networks of sites
that do not embody large spatial distances, the pair approximation can
work and the swapping test does not change measures of evolutionary
dynamics.  However, such networks do not capture important properties
of spatial heterogeneity.

As one of the key properties of spatial extent is the propagation of
organisms from one part of the space to the other over long distances,
we show that important insights can be gained by considering models of
percolation.  Percolation describes the physical propagation of,
\emph{e.g.,} fluids through a random medium.  In certain limits the
evolutionary behavior of spatial systems can be mapped onto
percolation behavior, demonstrating that investigations of such
systems which go beyond mean-field or scaling studies are relevant to
evolutionary dynamics.  This and other advances that go beyond the
mean field are necessary to fully describe spatial evolutionary
dynamics as they are necessary for the description of many physical
systems of spatial extent.  The complexities of spatially extended
evolutionary dynamical systems beyond the prototypical problem of
percolation create new demands and opportunities for advancing our
insight into the dynamics of heterogenous systems and their
implications for evolution.

\section{Model and Methods}
\label{sec:methods}
We make the issue of spatial heterogeneity concrete by focusing on a
specific model of ecological and evolutionary interest.  We take a
model of hosts and consumers interacting on a 2D spatial lattice.
Each lattice site can be empty (0), occupied by a host ($H$) or
occupied by a consumer ($C$).  We use the term {\em consumer} as a
general label to encompass parasites, pathogens and predators.  Where
convenient for examples, we will specialize to one or another of these
terminologies.  Hosts reproduce into adjacent empty sites with some
probability $g$ per site, taken as a constant for all hosts.
Consumers reproduce into adjacent sites occupied by hosts, with
probability $\tau$ per host; sometimes $\tau$ is fixed for all
consumers, but we also consider cases in which it is a mutable
parameter passed from parent to offspring.  We will refer to~$\tau$ as
the \emph{transmissibility.}  Hosts do not die of natural causes,
while consumers perish with probability $v$ per unit time (leaving
empty sites behind).  Because consumers can only reproduce into sites
where hosts live, the effective graph topology of reproductively
available sites experienced by the consumers is constantly changing
due to their very presence.  This makes the ecosystem an
\emph{adaptive network,} a system in which the dynamics {\em of\,} a
network and the dynamics {\em on} that network can occur at comparable
timescales and reciprocally affect one another~\cite{gross2008,
  gross2009, graeser2011, demirel2012}.  In this model, dynamics can
be highly complex, including spatial cascades of host and consumer
reproduction.  Even when a quasi-steady-state behavior emerges, as we
shall see, it is a consequence of fluctuations over extended space and
time intervals.

Several different types of biological interactions can be treated by
this modeling framework.  Hosts could represent regions inhabited by
autotrophs alone, while consumers represent regions containing a
mixture of autotrophs and the heterotrophs which predate upon
them~\cite{werfel2004}.  Alternatively, host agents could represent
healthy organisms, while consumers represent organisms infected with a
parasite or pathogen.  Thus, host--consumer models are closely related
to Susceptible--Infected--Recovered (SIR) models, which are
epidemiological models used to understand the spread of a disease
through a population.  SIR models describe scenarios in which each
individual in a network is either susceptible (S) to a pathogen,
infected (I) with it, or recovered (R) from it; susceptible nodes can
catch the disease from infected neighbors, becoming infected
themselves, while nodes which have become infected can recover from
the disease and are then resistant against further
infection. Susceptible, infected and recovered individuals roughly
correspond to hosts, consumers, and empty cells, respectively.  An
important difference between host--consumer models and epidemiological
models concerns the issue of {\em reinfection.}  In the host--consumer
model, an empty site left behind by a dead consumer can be reoccupied
by another consumer, but only if a host reproduces into it first.
Other research has considered models where R[ecovered] individuals can
also become I[nfected], with a different (typically lower) probability
than S[usceptible] ones, thereby incorporating imperfect immunity into
the model~\cite{jimenez2003,henkel2008}.  The degree of immunity is
independent of geography and the environment of the R[ecovered]
individual, unlike reoccupation in the host--consumer model.  Another
application is illustrated by the Amazon molly, {\em Poecilia
  formosa,} which is a parthenogenetic species: {\em P.~formosa,} all
of which are female, reproduce asexually but require the presence of
sperm to carry out egg development.  (This kind of sperm-dependent
parthenogenesis is also known as gynogenesis.)  {\em P.~formosa} are
thus dependent on males of other species in the same genus---usually
{\em P.~mexicana} or {\em P.~latipinna}---for reproduction.  Because
{\em P.~formosa} do not incur the cost of sex, they can outcompete the
species on which they rely, thereby possibly depleting the resource
they require for survival, {\em i.e.,} male fish~\cite{kokko2008,
  kokko2011}.  Thus, hosts could be regions containing sexual
organisms, with consumers standing for areas containing both sexual
and asexual individuals~\cite{kokko2008}.

\begin{figure}[h]
\includegraphics[width=4cm]{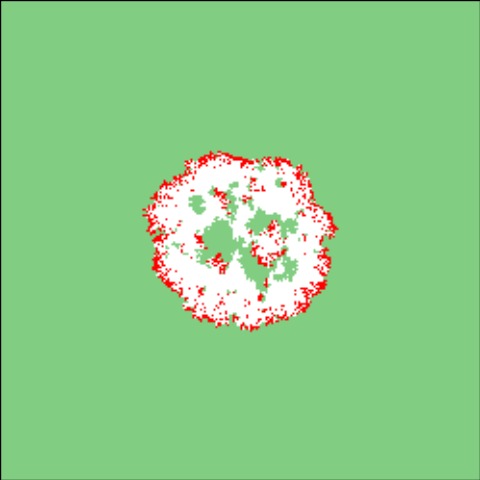}
\includegraphics[width=4cm]{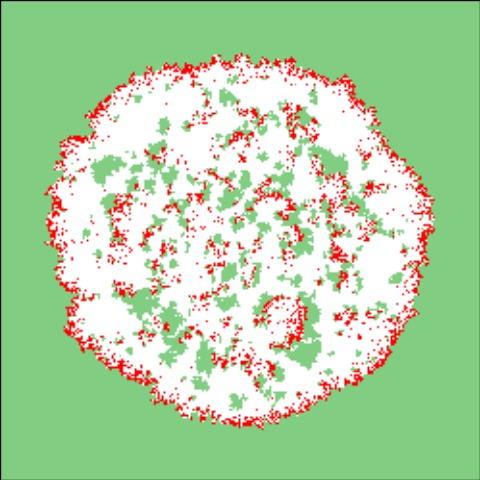}
\includegraphics[width=4cm]{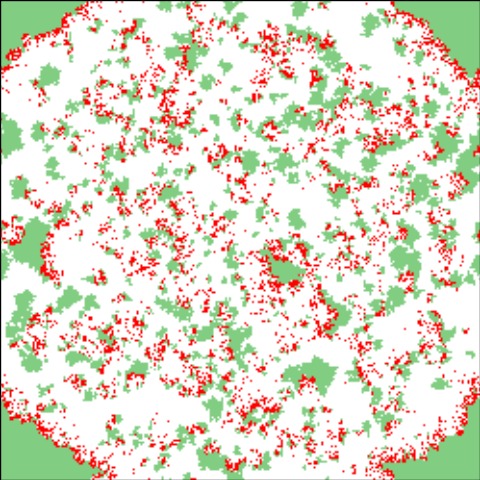}
\includegraphics[width=4cm]{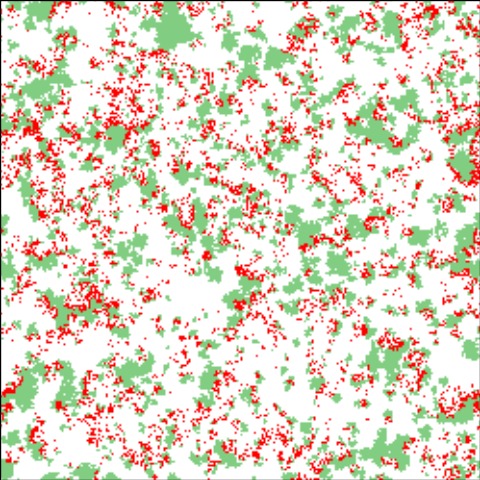}
\caption{\label{fig:animation} Snapshots of a simulated host--consumer
  ecosystem on a $250\times 250$ lattice, taken at intervals of 100
  generations.  Consumers are dark gray (red online), hosts are light
  gray (green online) and empty space is left white.  The simulation
  began with a single consumer at the center of the lattice, which
  gave rise to an expanding front of consumers.  The first image in
  this sequence shows the state of the ecosystem 100 generations into
  the simulation.  Hosts which survive the consumer wave recolonize
  the empty sites, leading to pattern formation.  Here, the host
  growth rate is $g = 0.1$, the consumer death rate is $v = 0.2$ and
  the consumer transmissibility is fixed at $\tau = 0.33$.}
\end{figure}

This host--consumer model displays waves of colonization, consumption
and repopulation.  Hosts reproduce into empty sites, and waves of
consumers follow, creating new empty regions open for host
colonization.  Therefore, clusters of hosts arise
dynamically~\cite{haraguchi2000, sayama2002, aguiar2003, sayama2003},
a type of pattern formation which can separate regions of the
resources available to pathogens into patches without the need for
such separation to be inserted manually.  Figure~\ref{fig:animation}
illustrates a typical example of this effect.  This is a specific
example of the general phenomenon of pattern formation in
nonequilibrium systems~\cite{sayama2000}.  Consumers are {\em
  ecosystem engineers}~\cite{jones1994, strayer2006, post2009,
  pringle2010, allen2013} which shape their local environment: an
excessively voracious lineage of consumers can deplete the available
resources in its vicinity, causing that lineage to suffer a Malthusian
catastrophe~\cite{haraguchi2000, cronin2005, kerr2006, kokko2008,
  messinger2009, lion2010, messinger2012, reigada2012}.  Because the
ecology is spatially extended, this catastrophe is a local niche
annihilation, rather than a global collapse~\cite{rauch2006}.  A
mutant strain with a high transmissibility can successfully invade in
the short term but suffer resource depletion in the medium term,
meaning that in a population where consumer transmissibilities evolve,
averages taken over long numbers of generations yield a moderate
value~\cite{goodnight2008, heilmann2010}.  This implies that an
empirical payoff matrix or reproduction ratio will exhibit nontrivial
timescale dependence~\cite{rauch2002, rauch2003, goodnight2008}.

This model is distinct from another approach to studying evolutionary
dynamics in spatial contexts, that of evolutionary game theory.
Game-theoretic models of spatially structured populations have been
explored at great length.  These investigations have found that
breakdowns of mean-field approximations are commonplace.  However,
evolutionary game theory has its own simplifying assumptions.  The
vast majority of studies consider only two-player games.  Population
size is usually taken to be constant, and population structure is
typically fixed in place.  In game-theoretic models, the benefits and
costs of different organism behavioral traits are parameters whose
values are chosen by the modeler.  By contrast, ``benefits'' and
``costs'' in host--consumer models are emergent properties which
depend on interactions over many generations.  Population size is not
fixed, and population structure is dynamical: the environment in which
different consumer varieties compete changes stochastically, in ways
affected by their presence.

\section{Results}

\label{sec:results}
\subsection{Evolution of Transmissibility}
\begin{figure}[h]
\includegraphics[width=7.5cm]{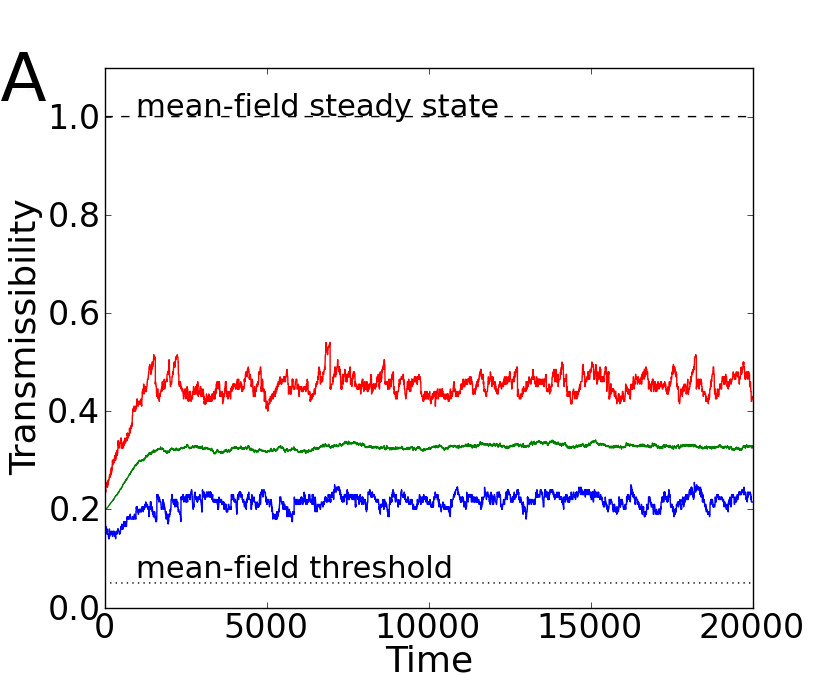}
\includegraphics[width=8cm]{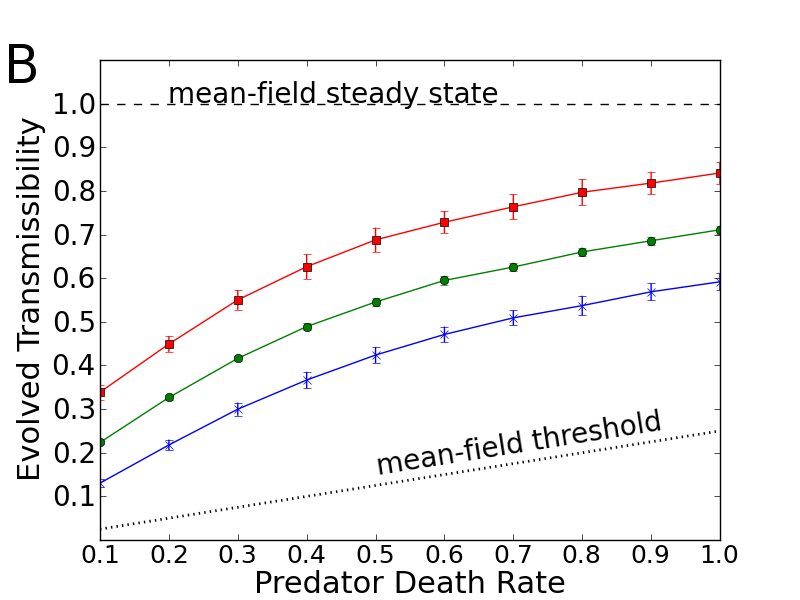}
\caption{\label{fig:minmax-6apr2011} {\bf (A)} Minimum (blue), average
  (green) and maximum (red) transmissibility $\tau$ for a consumer
  population over time, with~$g = 0.1$ and $v = 0.2$.  (The mutation
  rate is $\mu = 0.255$ and the step size is $\Delta\tau = 0.005$, as
  was used in reference~\cite{werfel2004}.)  The average $\tau$ tends
  to a quasi-steady-state value dependent on~$g$ and $v$; if the
  simulation is started with $\tau$ set to below this value, the
  average $\tau$ will increase, and likewise, the average $\tau$ will
  decrease if the consumer population is initialized with $\tau$ over
  the quasi-steady-state value~\cite{werfel2004}.  The horizontal
  dotted line indicates the threshold value of~$\tau$ which, in a
  mean-field model, is the smallest value at which a consumer
  population can sustain its numbers.  The dashed line indicates the
  value to which $\tau$ would trend in a well-mixed ecosystem.  {\bf
    (B)} Minimum, average and maximum $\tau$ as a function of~$v$,
  with~$g = 0.1$.  The dotted line shows the minimum sustainable
  $\tau$ as predicted by mean-field approximation.  Each point is
  found by averaging over 15,000 timesteps.  Error bars indicate one
  standard deviation.}
\end{figure}

We investigate evolution in the spatial host–consumer ecosystem
through simulation and analytic discussion.  If the transmissibility
$\tau$ is made a heritable trait, passed from a consumer to its
offspring with some chance of mutation, what effect will natural
selection have on the consumer population?
Figure~\ref{fig:minmax-6apr2011}(A) shows the average, minimum and
maximum values of the transmissibility $\tau$ observed in a population
over time.  The average $\tau$ tends to a quasi-steady-state value
dependent on the host growth rate $g$ and the consumer death rate $v$;
if the simulation is started with $\tau$ set to below this value, the
average $\tau$ will increase, and likewise, the average $\tau$ will
decrease if the consumer population is initialized with $\tau$ over
the quasi-steady-state value.  Even when the average $\tau$ has
achieved its quasi-steady-state value, the population displays a wide
spread of transmissibilities whose extremes fluctuate over
time~\cite{werfel2004}.

In a well-mixed ecosystem, the average $\tau$ of the population will
tend to 1, maximizing the reproductive rate of the individual
consumer.  This occurs because each consumer on average experiences
the same environment as any other, and thus has the same number of
hosts available to reproduce into.  A consumer with a higher $\tau$ has
a higher reproduction rate and therefore evolutionary dominance up to
the highest possible value, 1.  The observation of a
quasi-steady-state value below 1 is an important result.  This is the
first breakdown of the mean-field approximation, and it indicates the
inapplicability of traditional assumptions about fitness optimization,
with implications for the origins of reproductive restraint,
communication-based altruism and social behaviors in
general~\cite{rauch2002, rauch2003, werfel2004, aguiar2004, rauch2006,
  goodnight2008}.

One can avoid $\tau$ tending to 1 in a panmictic system by imposing
some extra constraint, such as a tradeoff between transmissibility and
lethality, where higher transmissibility becomes impossible due to
lethality that prevents transmission.  This tradeoff between
infectiousness and lethality can be considered as a within-host
version of resource overexploitation that here occurs at the
population level.  Such within-host tradeoffs are difficult to
establish empirically in living populations~\cite{froissart2010,
  asplen2012}.  Often, one lacks pertinent information, such as the
functional relationship between pathogen load and disease transmission
probability, or the extent to which empirical proxies for pathogen
load predict actual host mortality~\cite{hawley2013}.  An empirical
observation of low virulence should not by itself be taken as evidence
that a tradeoff exists: it may well be that another condition, such as
panmixia, fails to obtain.  The behavior of spatial models makes clear
that the relevant scale of the limiting factor is not necessarily
within the individual host.

Another difference between spatial and nonspatial host--consumer
systems is the rate at which consumers must reproduce in order to
sustain their population.  One can calculate the minimum sustainable
value of~$\tau$ in the mean-field approximation~\cite{mobilia2006b} by
balancing the birth and death rates.  If the host population is small
compared to the total ecosystem size, then the minimum sustainable
$\tau$ is the value which satisfies $k\tau = v$, where $k$ is the
number of neighbors adjacent to a site.  For the parameters used in
Figure~\ref{fig:minmax-6apr2011}(A), this value would be 0.05, which
is substantially smaller---by a factor of~4---than the lowest $\tau$
seen in the evolving spatial population.  Consumer populations
with~$\tau$ at the mean-field threshold are not sustainable in the
spatial case.  This is easily verified by numerical simulations or by
using the mean-field equations for the host--consumer
dynamics~\cite{aguiar2003b, aguiar2004, aguiar2003b-errata}.
Stochastic fluctuations suppress the active phase, \emph{i.e.,} the
range of parameter values which permit a living consumer population is
reduced~\cite{mobilia2006b}.

\subsection{Timescale Dependence of Invasion Success}
\label{sec:timescale}
\begin{figure}[h]
\includegraphics[width=8cm]{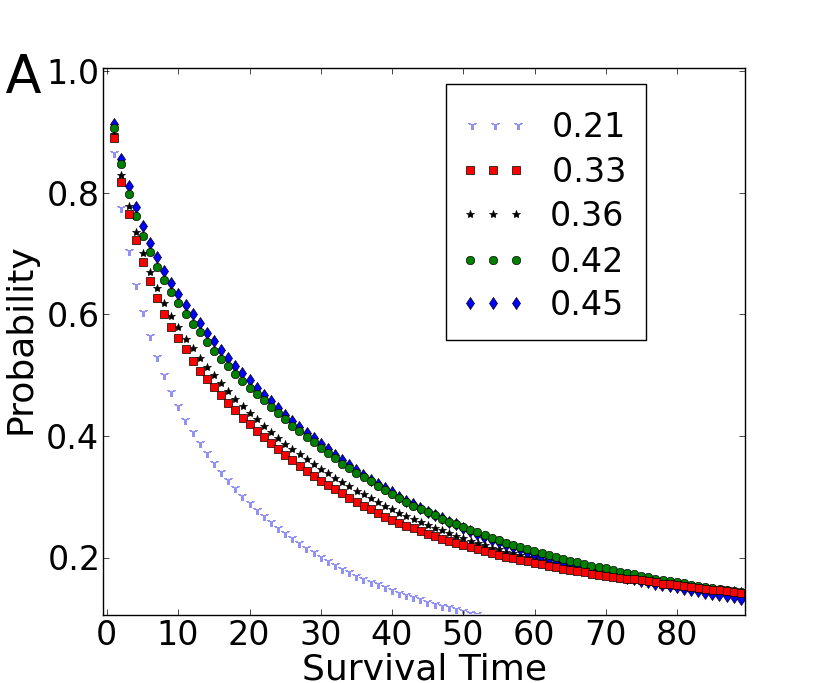}
\includegraphics[width=8cm]{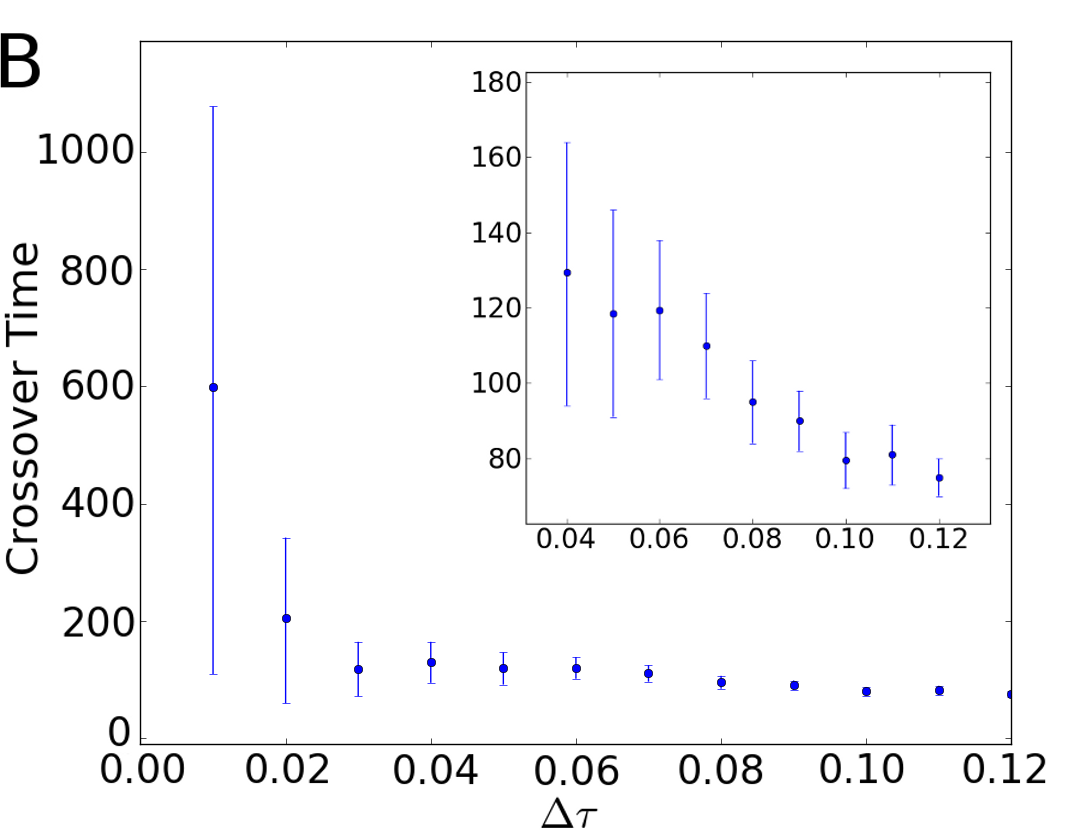}
\caption{\label{fig:survival-time-comparison} {\bf (A)} Survival
  probability as a function of time for five scenarios: injecting
  mutants with the same transmissibility as the native consumers,
  three examples of injecting mutants with transmissibility higher
  than the $\tau$ of the native consumers, and an example of injecting
  mutants with lower $\tau$ than the native population. {\bf (B)} Time
  intervals during which the survival-probability curves for the
  native and invasive strains overlap.  $\Delta\tau$ indicates the
  difference between the invasive and native transmissibilities.  The
  closer the mutant trait value is to the resident, the greater the
  duration of time over which the survival-probability curves for the
  native and mutant strains overlap.  Here, overlap is defined by
  probabilities being coincident at the 95\% confidence level; using
  other overlap criteria gives qualitatively the same results.  Inset:
  magnified view of the $\Delta\tau \geq 0.04$ region.}
\end{figure}

A key question about an ecological system is whether a new variety of
organism, having a different genetic character and phenotypic trait
values, can successfully invade a native population.  If a mutant
consumer strain with fixed transmissibility $\tau_m$ can successfully
invade a population of transmissibility $\tau_0 < \tau_m$, then we
expect the time-averaged value of~$\tau$ seen in the evolving system
to be larger than $\tau_0$.  To investigate this, we simulate
scenarios where the native population has $\tau$ close to the average
value seen in the evolutionary case described above.  We then inject a
mutant consumer strain with significantly larger $\tau$ and study the
results.  For a typical example, we see from
Figure~\ref{fig:minmax-6apr2011}(A) that when $g = 0.1$ and $v = 0.2$,
the average $\tau$ is approximately 0.33.  So, we simulate $\tau_m =
0.45$ mutants entering an ecosystem whose native population has
$\tau_0 = 0.33$.  Initially, the mutants prosper, but they ultimately
fail to invade.  As shown in
Figure~\ref{fig:survival-time-comparison}, the probability of a
$\tau_m = 0.45$ strain surviving for tens of generations after
injection is larger than that of a $\tau_0 = 0.33$ strain. That is,
mutants with the higher $\tau$ can out-compete the neutral case.
However, after $\approx\!74$ generations, the survival-probability
curves cross.  Observed over longer timescales, the mutant strain is
less successful than the native variety.  This pattern is consistent
for $\tau_m > \tau_0$: the average transmissibility seen in the
evolutionary case stands up to invasive varieties.  This key result
manifests the distinctive properties of the spatial structure of the
model.  The underlying reason for this result is that the mutants
encounter the resource limitations imposed by the patchy native
population.  Over short timescales, the mutant strain enjoys the
resources available within the local patch, consuming those resources
more rapidly than can be sustained once it encounters the limitations
of the local patch size.  In this way, the initial generations of the
mutant strain ``shade'' their descendants.  Thanks to
descendant-shading, short-term prosperity is not a guarantee of
medium- or long-term success.

This is to be contrasted with what happens in a well-mixed ecosystem.
In the well-mixed scenario, consumer strains with higher $\tau$
successfully invade and displace the native population with a high
probability.  The invasion success is consistent with the dynamics of
a continuously evolving ecosystem.  If $\tau$ is made an evolvable
trait in simulated panmictic systems, the average $\tau$ of the
population will tend to~1, as predicted by the mean-field analytic
proof.  There is no difference in a well-mixed scenario between
short-term and long-term success.  Descendant-shading does not occur
in the well-mixed case.  This follows from the lack of distinction
between local patches and large-scale structure.

One common measure of evolutionary success is the expected relative
growth rate of the number of offspring of a mutant individual within a
native population, \emph{i.e.,} the relative growth rate of a mutant
strain.  This rate, known as the {\em invasion fitness,} is often used
to investigate the stability of an evolutionary
ecosystem~\cite{vanbaalen1998, vanbaalen2000, lion2008}.  If the
invasion fitness is found to be positive, the native variety is judged
to be vulnerable to invasion by the mutant.  Conversely, if the
invasion fitness is found to be negative, the native variety is deemed
to be stable.  For the spatial host--consumer ecosystem, this method
gives qualitatively incorrect predictions for evolutionary dynamics.

Our investigation builds on earlier work which studied the timescale
dependence of fitness indicators in spatial host--consumer
ecosystems~\cite{rauch2002, rauch2003}. In this paper we have
augmented the prior work by considering the survival probability to
show the effects of varying $\tau$.  We have also more systematically
shown the number of generations until dominance of the evolutionary
stable strain.  In addition, we reported the case of a mutant strain
invading a background population, clarifying the conceptual and
quantitative results of those earlier works, which considered instead
scenarios complicated by multiple ongoing mutations.

\subsection{Pair Approximations}

The inadequacy of mean-field treatments of spatial systems motivates
the development of more elaborate mathematical methods.  In this
section, we review one such methodology, based on augmenting
mean-field approximations with successively higher-order correlations,
and we test its applicability to our host--consumer spatial model.
The numerical variables used in this methodology are probabilities
which encode the state of the ecosystem and can change over time.  One
such variable is, for example, the probability $p_a$ that a lattice
site chosen at random contains an organism of type $a$.  Another is
$p_{ab}$, the probability that a randomly-chosen {\em pair} of
neighboring sites will have one member of type $a$ and the other
of type $b$.  The change of these quantities over time is usually
described by differential equations, for which analysis tools from
nonlinear dynamics are available~\cite{vanbaalen2000, haraguchi2000,
  aguiar2004, lion2008, rozhnova2009, allen2010, araujo2010}.

The importance of the joint probabilities $p_{ab}$ is that they
reflect correlations which mean-field approximations neglect.  To
understand the relevance of the joint probabilities $p_{ab}$, consider
a scenario where an invasive mutant variety forms a spatial cluster
near its point of entry.  Let $p_M$ be the probability that a lattice
site chosen at random contains a mutant-type organism, and let
$p_{MM}$ denote the probability that a pair of neighboring sites
chosen at random will both be occupied by mutant-type organisms.  Then
the average density of invasive mutants in the ecosystem, $p_M$, will
be low, while the conditional probability that a neighbor of an
invasive individual will also be of the invasive type, $q_{M|M} =
p_{MM} / p_{M}$, will be significantly higher.  (It is typical in
theoretical spatial ecology to denote conditional probabilities
with~$q$, rather than $p$~\cite{tgoei2000}.)  A discrepancy between
the conditional probability $q_{a|b}$ and the overall probability
$p_a$ can persist when the ecosystem has settled into a
quasi-steady-state behavior, and is then an indicator of spatial
pattern formation.  

\begin{figure}[h]
\includegraphics[width=8cm]{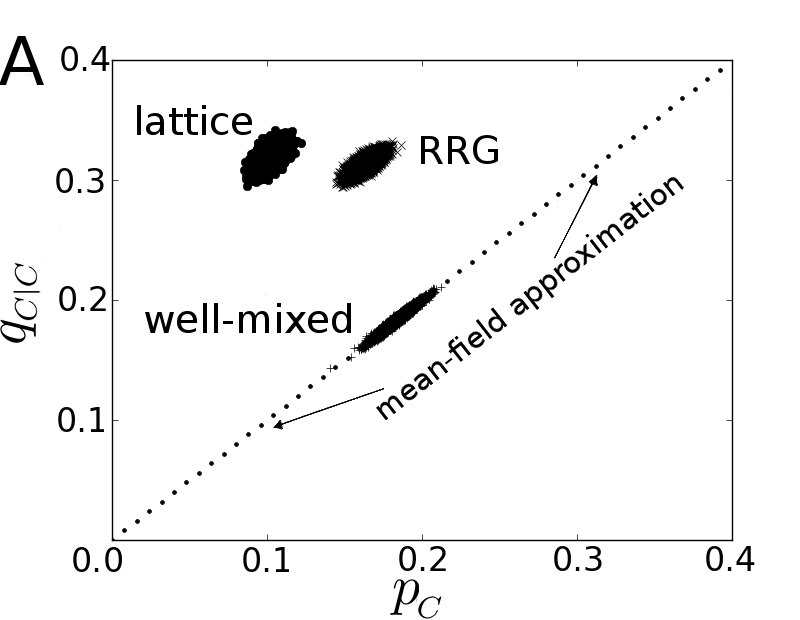}
\includegraphics[width=8cm]{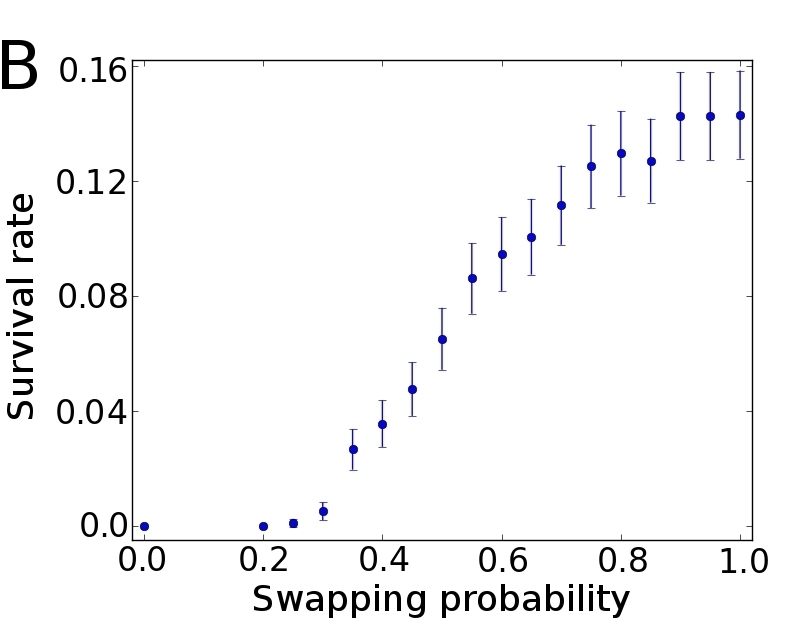}
\caption{\label{fig:qcc-over-pc} {\bf (A)} Pairwise conditional
  probability $q_{C|C}$ plotted against the average density of
  consumers, $p_C$, for three variations on the host--consumer model:
  a well-mixed case in which mean-field theory is applicable, a random
  regular graph (in which each site has exactly four neighbors) and a
  2D square lattice.  The dotted line, $p_C = q_{C|C}$, indicates the
  mean-field approximation.  $10^4$ timesteps were computed for each
  case.  The well-mixed case is simulated by dynamically rewiring
  sites at each time step, precluding the generation of spatial
  heterogeneity; consequently, the pairwise correlation $q_{C|C}$ is
  within statistical variation equal to~$p_C$ ($R^2 = 0.953$).  A
  random regular graph (RRG) with random but static connections does
  develop spatial heterogeneity so that $q_{C|C}$ is not the same
  as~$p_C$ ($R^2 = 0.581$). The discrepancy is even stronger in the
  lattice case ($R^2 = 0.304$).  {\bf (B)} Success rate of invasive
  mutant strains as a function of swapping probability.  Voracious
  mutant strains with $\tau = 0.45$ are introduced into a lattice
  ecosystem defined by a host growth rate of $g = 0.1$, a consumer
  death rate $v = 0.2$ (the same for both consumer varieties), and a
  native consumer transmissibility of $\tau = 0.33$.  Average success
  rates are found by simulating 2000 invasions per value of the
  swapping probability parameter; error bars indicate 95\%-confidence
  intervals.  Increasing the fraction of possible swaps which are
  actually performed makes the voracious invasive strain more likely
  to take over the ecosystem.}
\end{figure}

Applying this idea to the spatial host--consumer model, let $p_C$ be
the probability that a lattice site chosen at random contains a
consumer, and let $q_{C|C}$ denote the conditional probability that
lattice site adjacent to a consumer will also be occupied by a
consumer.  Figure~\ref{fig:qcc-over-pc}(A) shows $p_C$ and $q_{C|C}$
measured during the course of numerical simulations.  In a well-mixed
scenario (where we expect the mean-field approximation to be
applicable), the average consumer density $p_C$ and the
consumer--consumer pairwise correlation $q_{C|C}$ are essentially
equal over time.  In the spatial lattice scenario, $p_C$ and $q_{C|C}$
are noticeably different.

Treating the correlations $q_{a|b}$ as not wholly determined by the
probabilities $p_a$ is a way of allowing spatial heterogeneity to
enter an analytical model.  Whether it is a {\em sufficient} extension
in any particular circumstance is not, {\em a priori,} obvious.
Typically, the differential equations for the pair probabilities
$p_{ab}$ depend on triplet probabilities $p_{abc}$, which depend upon
quadruplet probabilities and so forth.  The standard procedure is to
truncate this hierarchy at some level, a technique known as {\em
  moment closure}~\cite{matsuda1992, vanbaalen1998, rozhnova2009,
  do2009, demirel2012}.  Moment closures constitute a series of
approximations of increasing intricacy~\cite{buice2009,dodd2009}.  The
simplest moment closure is the mean field approximation; going beyond
the mean field to include second-order correlations but neglecting
correlations of third and higher order constitutes a {\em pair
  approximation.}  These approximations do not incorporate all of the
information about spatial structure which may be necessary to account
for real-world ecological effects~\cite{vanbaalen1998}.

\subsection{Organism Swapping}
\label{sec:swapping}
Several factors have been identified which undermine pair
approximations~\cite{vanbaalen1998, aguiar2003b, aguiar2004,
  aguiar2003b-errata, werfel2004, szabo2007, allen2010, allen2011,
  demirel2012, smaldino2013}.  In our model, we can directly test the
efficacy of pair approximations in a completely general way.  The key
idea is to transplant individuals in such a way that the variables
used in the moment-closure analytical treatment remain unchanged.  At
each timestep, we look through the ecosystem for isolated consumers,
that is, for individual consumers surrounded only by a specified
number of hosts and empty sites.  We can exchange these individuals
without affecting the pairwise correlations.  For example, if we find
a native-type consumer adjacent to three hosts and one empty site, we
can swap it with an invasive-type consumer also adjacent to three
hosts and one empty site.  We can also exchange isolated pairs of
consumers in the same way.  The variables used in the moment-closure
treatment remain the same.  Were the moment-closure treatment valid,
we would expect the dynamics to remain unchanged when we perform such
exchanges.

When we perform the simulation, however, swapping strongly affects the
dynamics.  With this type of swapping in effect, mutants with higher
$\tau$ can invade a native population with lower $\tau$.  In one
typical simultation run with a native $\tau$ of~0.33 and an invasive
$\tau$ of~0.45, the invasive strain succeeded in~1,425 of~10,000
injections.  Without swapping, the number of successful invasions is
{\em zero.}

Swapping can be considered as creating a new ecosystem model with the
same moment-closure treatment as that of the original.  The behavior
of invasive strains is different, because transplanting organisms
allows invasive varieties to evade localized Malthusian catastrophes.
Swapping opens the ecosystem up to invasive strains, since, in
essence, it removes individuals from the ``scene of the crimes''
committed by their ancestors.

This type of swapping is, to our knowledge, a new test of
moment-closure validity.  Randomized exchanges have been incorporated
into computational ecology simulations for different purposes.  For
example, research on dispersal rates in an island model shuffled
individuals in such a way that the population size of each island was
held constant~\cite{poethke2007}.

If, instead of performing every permissible swap, we transplant
organisms with some probability between 0 and 1, we can interpolate
between the limit of no swapping, where invasions always fail, and the
case where pair approximation is most applicable and invasions succeed
significantly often.  The results are shown in
Figure~\ref{fig:qcc-over-pc}(B) and indicate that the impact of
swapping becomes detectable at a probability of $\approx\!0.25$ and
effectively saturates at a probability of $\approx\!0.9$.  

Our swapping method allows us to test the significance of
complications which can undermine pair approximation techniques or
make them impractical to apply, several of which have been identified.
First, introducing \emph{mutation} into a game-theoretic dynamical
system can make pair approximation treatments of that system give
inaccurate predictions~\cite{allen2010, allen2011}.  Second, when the
evolving population has a network structure, the presence of
\emph{short loops} in the network often makes pair approximations
fail~\cite{szabo2007}.  For example, in a triangular lattice, one can
take a walk of three steps and return to one's starting point, whereas
on a hexagonal lattice, the shortest closed circuit is six steps long.
A pair approximation can work well for a dynamical system defined on
the hexagonal lattice but fail when the same dynamics are played out
on a triangular one.  This happens because the short loops provide
opportunities for contact which the coarse-graining necessary for a
pair approximation will miss.  This effect is amplified in adaptive
network models, where the underlying network changes dynamically in
response to the population living upon it.  In such cases, even
extending the moment closure to the triplet level brings little
improvement~\cite{demirel2012}.  Third, \emph{fluctuating population
  sizes} make pair approximations significantly more cumbersome to
construct, leading to systems of differential equations which are too
intricate to be significantly illuminating.  In a game-theoretic model
where a lattice is completely filled at all times with cooperators and
defectors, there is one independent population density variable and
three types of pairs.  By contrast, in an ecological model where two
consumer varieties are competing within an adaptive network of hosts,
a pair approximation requires nine independent
variables~\cite{aguiar2003b, aguiar2004, aguiar2003b-errata}.
Modeling phenomena of biological interest can easily increase the
complexity still more.  For example, if organism behavior changes in
response to social signals~\cite{werfel2004}, the number of possible
states per site, and thus the number of dynamical variables in a pair
approximation treatment, increases further.  Fourth, the
pair-approximation philosophy of averaging over all pairs in the
system impedes the incorporation of \emph{environmental
  heterogeneities,} including biologically crucial factors like
variable organism mobility, background toxicity or other localized
``costs of living,'' and resource availability~\cite{smaldino2013}.
Finally, \emph{dynamical pattern formation} creates spatial
arrangements which the pair approximation does not
describe~\cite{vanbaalen1998}.

\subsection{Effect of Substrate Topology}
\label{sec:substrate}

It is instructive to compare the spatial lattice ecosystem with the
host--consumer model defined on a random regular graph (RRG).  In an
RRG, each node has the same number of neighbors, as they do in a
lattice network, but the connections are otherwise random.  RRGs have
been used as approximations to incorporate the effects of spatial
extent into population models, as they make for more tractable
mathematical treatments, although they are typically less realistic
than spatial lattices~\cite{lion2009}.  The network structure is set
at the beginning of a simulation and does not change over time.  The
important aspect of this network as compared to the spatial case is
that there exist short paths of links that couple all nodes of the
network.  This is quite different from the spatial case, where strains
in one part of the network cannot reach another in only a few
generations due to the need to traverse large numbers of spatially
local links.

When we simulate our host--consumer ecosystem on an RRG, we find that
an invasive consumer strain with higher transmissibility $\tau$ can
out-compete and overwhelm a native consumer population with lower
$\tau$.  In one typical simulation run, using the native and invasive
$\tau$ values of 0.33 and 0.45 respectively, 2,233 out of 10,000
invasions were successful, whereas on the lattice {\em no} invasion
succeeded using the same parameters.  Thus, the RRG does not capture
the essential features of the spatial scenario.  In particular, our
results show that the RRG case is more like the well-mixed case than
the spatial lattice, as far as stability against invasion is
concerned.

Our swapping test provides insight into the utility of the pair
approximation, which can be effective for the RRG even though it is
not for the spatial case.  Consider the pairwise correlation value
$q_{C|C}$, which would be a variable for a pair approximation
treatment.  On an RRG, the underlying network topology provides enough
locality that $p_C$ and $q_{C|C}$ are unequal, distinct from the well
mixed case as shown in Figure~\ref{fig:qcc-over-pc}(A).  This means
that the pair approximation is nontrivial for the RRG as it
incorporates the difference between~$q_{C|C}$ and $p_C$, which would
not be contained in a mean-field treatment.  We can also implement
swapping on the RRG, where invasions can succeed without it; as
expected, swapping does not affect the success rate on the RRG.  With
10,000 simulated invasions for each case, the 95\%-confidence interval
for the difference in success rates between full swapping and none is
$0.004 \pm 0.01$.  Thus, the pair approximation may be successful in
this network topology.  However, this does not mean that the RRG or
the pair approximation capture the full significance of a spatial
system, because the RRG network does not embody essential properties
of spatial extent---separation by potentially large distances.

\subsection{Percolation}
\label{sec:percolation}
In order to obtain quantitatively or even qualitatively correct
predictions for spatial host--consumer evolutionary dynamics,
different approaches are needed.  Having encountered the limitations
of moment closures, we now demonstrate a change of perspective which
yields quantitatively useful results.  In certain situations, the
process of pathogen propagation through the host population
distributed in space can be mapped onto a {\em percolation} problem. A
topic widely investigated in mathematics, percolation theory deals
with movement though a matrix of randomly placed obstacles. A
prototypical percolation problem is a fluid flowing downhill through a
regular lattice of channels, with some of the lattice junction points
blocked at random. The key parameter is the fraction of blocked
junction points. If this fraction is larger than a certain threshold
value, the fluid will be contained in a limited part of the
system. However, if the blocking fraction is below the threshold, the
fluid can percolate arbitrarily far from its starting point. This is a
\emph{phase transition,} a shift from one regime of behavior to
another, in this case between a phase in which fluid flow can continue
indefinitely and one in which flow always halts. Similar issues arise
when a pathogen propagates by cross-infection through a set of
spatially arranged hosts. Sufficiently many hosts in mutual contact
are required for the pathogen to propagate successfully. Pathogen
strains therefore survive or die out over time depending on whether
percolation is or is not possible~\cite{mobilia2006b, arashiro2006,
  davis2008, salkeld2010, givan2011, neri2011}.

One important goal of studying host--pathogen models is knowing the
pathogen properties that enable its survival in a population, or
equivalently what prevents it from persisting in a population. The
growth rate of a pathogen in a population can be an important public
health concern. We therefore focus on analyzing the minimum value of
the transmissibility that enables a pathogen population to persist,
and the growth dynamics of population sizes near that transition.

Of essential importance to the quantitative theoretical and empirical
analysis is the recognition that infected population growth can be
described by power laws $n \sim t^z$, with an exponent that differs from
that of the mean field. Identifying the value of the power $z$ is
important to practical projections of the number of infected
individuals. The initial growth curve of infected populations can be
correctly extrapolated if the exponent is known, guiding public health
responses.  Knowing what impediments are needed to prevent further
propagation can even better guide public health intervention
strategies.

\begin{figure}[h]
\includegraphics[width=8cm,height=6.4cm]{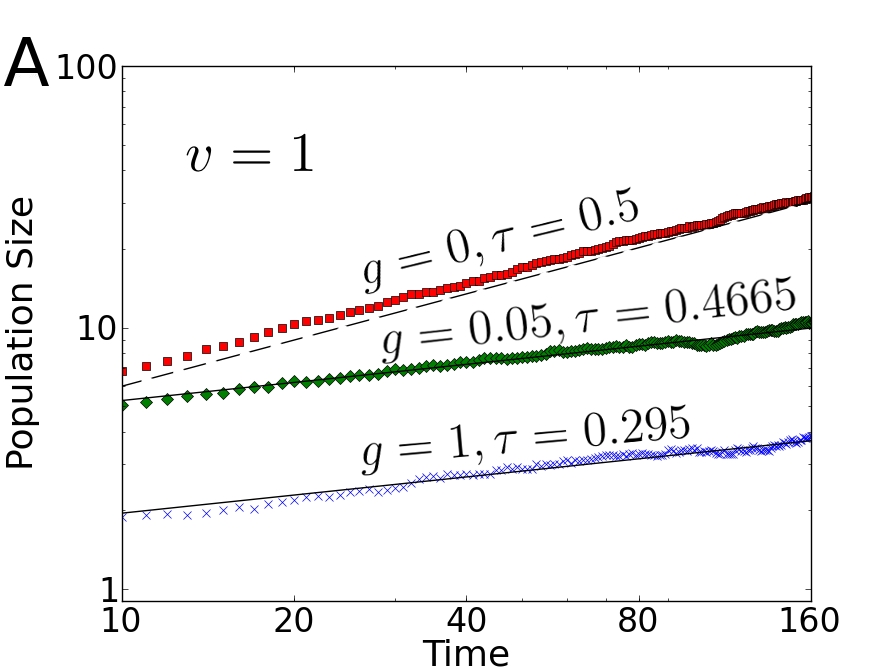}
\includegraphics[width=8cm]{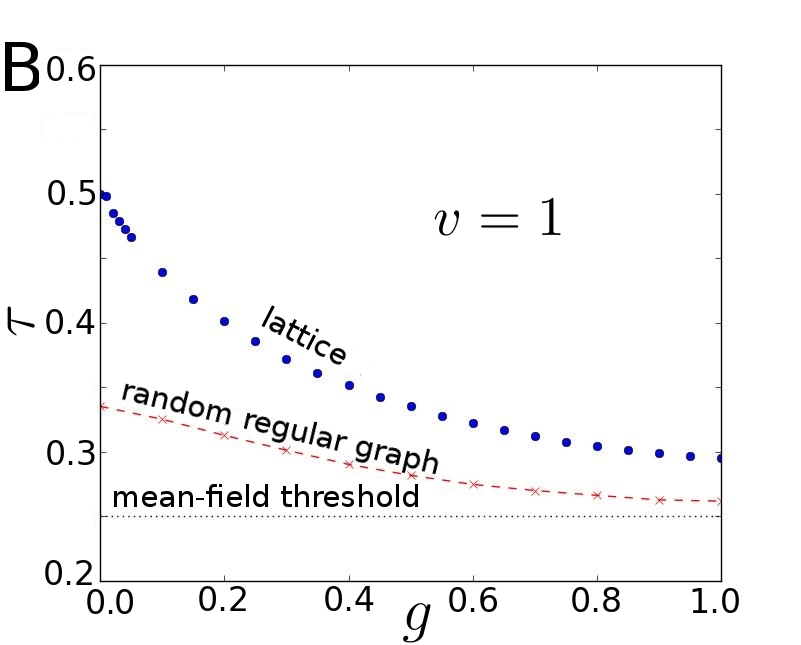}
\caption{\label{fig:percolation} {\bf (A)} Population size as a
  function of time, averaged over $10^3$ simulation runs, for $\tau$
  values near the transition points at $g = 0$ and $g = 0.05$ on the
  spatial lattice, with $v = 1$.  Dashed and solid lines indicate the
  population growth for systems at dynamic percolation and directed
  percolation transitions respectively, showing that these transitions
  have the characteristic properties of those universality classes.
  {\bf (B)} Critical $\tau$ for the host--consumer ecosystem with $v =
  1$.  The transition line crosses over from the dynamic percolation
  universality class at $g = 0$ to directed percolation between $g =
  0.015$ and $g = 0.02$.  Red Xs indicate the transition curve for the
  host-consumer dynamics on a random regular graph (RRG) of
  uniform degree 4; the dashed line connecting them is to guide the
  eye.  The RRG transition is neither directed percolation nor dynamic
  percolation.  }
\end{figure}

We show in Figure~\ref{fig:percolation} the results of numerical
simulations which indicates that the consumer extinction transition,
when the transmissibility $\tau$ becomes just large enough that the
consumer population sustains itself, lies in the {\em directed
  percolation universality class}~\cite{hinrichsen2000, henkel2008,
  saif2010, wendykier2011, szolnoki2011}.  A similar result has been
reported for related models~\cite{mobilia2006b, arashiro2006},
consistent with those models being in the same universality class.
The directed percolation universality class is a large set of models,
all of which exhibit a phase transition between two regimes of
behavior, and all of which behave in essentially the same way near
their respective transition points.  The scenario of fluid flow
through a random medium considered above is a classic example of a
directed percolation-class model, but many others exist as
well~\cite{hinrichsen2000, henkel2008}.  The {\em critical exponents}
describe how properties of the modeled system vary over time or as a
function of how far the control parameter is from the critical point.
They are the same for all systems in the universality class.  Other
universality classes exist as well, with different classes having
different quantitative values for the critical exponents.  Identifying
the universality class a system belongs to enables us to study a
complicated phenomenon by examining a simpler representative of its
class instead.  This is convenient, because regions of parameter space
near phase transitions are precisely where mean-field and
moment-closure approximations are least reliable, even for
short-timescale modeling.  Near the phase transition, stochastic
fluctuations create dynamical patterns with a wide range of sizes.  In
Figure~\ref{fig:percolation}(A), we see that percolation theory gives
quantitatively correct predictions for the growth of consumer
population sizes in the spatial host--consumer model.

We can understand the $g = 0$ and $g = 1$ extremes by mapping the
host--consumer model onto other stochastic models for which exact or
approximate results are available.  When $g = 0$, the host--consumer
model maps onto the SIR epidemic process~\cite{henkel2008}.  In turn
the SIR model on the square lattice can be understood in terms of bond
percolation on the square lattice~\cite{grassberger1997}, for which
the transition point is known exactly~\cite{kesten1980, grimmett1999}.
We can therefore predict analytically that the critical $\tau$ on the
square lattice is~0.5.  Percolation theory also gives a prediction for
the critical $\tau$ on an RRG: it should be approximately
$1/3$~\cite{newman2002}.  These both match the simulation results seen
in Figure~\ref{fig:percolation}(B).

In contrast, when $g = 1$, empty sites are filled as quickly as
possible, so the behavior of the host--consumer model should resemble
that of an epidemic model with only Susceptible and Infected sites.
In this case, the transition point of the epidemic model on the square
lattice is only known numerically~\cite{grassberger1997}.  The
numerical value, $\approx\!0.29$, does agree with the critical $\tau$
found by simulating the host--consumer model at~$g = 1$.

Thus, in the limiting cases of $g = 0$ and $g = 1$, the host--consumer
model is roughly equivalent to the SIR and SIS epidemic models.
However, a host--consumer model with~$0 < g < 1$ has dynamical
behavior distinct from an epidemic model which allows reinfection of
Recovered sites.  The key difference is that reoccupying an empty site
with a consumer requires prior recolonization by a host, whereas the
vulnerability of a R[ecovered] individual to becoming I[nfected] is
defined as an intrinsic property of the R[ecovered] type.  This
changes the role of ecology: both models incorporate space, but the
effect of spatial extent is different.  This manifests as a change in
the shape of the critical-threshold curve, as well as a change in
universality class~\cite{grassberger1997, henkel2008}.

Furthermore, when we use an RRG topology instead, comparing
host--consumer dynamics at~$g = 1$ with an SIS epidemic model reveals
their transitions to take place at different thresholds.  For the
host-consumer model, the critical $\tau$ on an RRG is approximately
0.2615, while the SIS threshold is approximately 0.34~\cite{gleeson2011}.

The analysis of percolation behavior near the transition point from
physics, maps directly onto the critical public health problem of the
growth of infected populations, and more generally onto the dynamics
of evolutionary systems.  For these systems the mean field treatment
fails and the standard transmission of infectious diseases in a
population need not apply.  Applications to real world systems must
accommodate the actual network of connectivity. This network can also
be modified by intervention strategies.

\section{Discussion}
\label{sec:discussion}

Understanding the effects of spatial extent is a vital part of
evolutionary ecology.  Spatial extent changes the quantitative and
qualitative characteristics of a model's evolutionary behavior,
compared to well-mixed models.  The short-term success rate of novel
genetic varieties is not indicative of their long-term chance of
success relative to the prevalent type.  Standard stability criteria
fail to reflect the actual stability achieved over time.  We must
instead consider extended timescales because they are determined by
spatial patterns, whose ongoing formation is an intrinsic part of
nonequilibrium evolutionary dynamics.  Our analysis provides a clear
understanding of why there are dramatic differences between spatial
models and mean-field models, which simplify away heterogeneity
through mixing populations, averaging over variations or mandating a
globally connected patch structure.  We have further shown that
transplanting organisms dramatically changes the dynamics of spatial
systems, even when we preserve local correlations as would be
considered in a pair approximation treatment.  Our results prove that
any model striving to capture the effects of heterogeneity that does
not change its behavior with organism transplanting cannot fully
capture the dynamics of spatial evolution.  The following subsections
summarize the general conclusions we draw from these results.

\subsection{Defining Fitness}

In our host--consumer models, each individual either survives or it
does not, and any individual has a specific number of offspring and
survives over a certain amount of time; that is to say, an
``individual fitness'' (in the terminology of~\cite{orr2009}) is a
well-defined concept.  To find {\em expected} individual fitness, or
{\em average} individual fitness, we must define a set of individual
organisms over which to take an average, which is the very concept we
have established to be problematic.  Consequently, derived notions of
fitness, which depend on comparisons between such
averages~\cite{orr2009}, become elusive, context-dependent quantities.
The problem is both temporal and spatial: Average relative fitness in
one generation is not necessarily a good measure of the long-term
success of a strain in one, or a combination of, the broad variety of
dynamically-generated niches.  This problem is not, however, the same
as the traditional concept of variation of fitness across a static set
of niches, because the niche dynamics of our spatially explicit model
ensures that evolutionary outcomes are not reflected in any standard
definition of the average.

One might be tempted to call the behavior of voracious invasive
consumers ``frequency-dependent fitness''~\cite{page2002,hartl2007},
as the invasive strain is successful initially when rare but fails
when it becomes more common. The term ``frequency-dependent fitness''
is, however, a misnomer in this context, because the organism type is
rare and successful when it is newly introduced, but as it declines to
extinction it becomes rare and unsuccessful.  Nor can we attribute the
decline to the frequency of hosts: the average population density of
hosts remains essentially unchanged, because the boom and the
following bust are localized.  {\em Frequency,} being defined by an
average over the whole ecosystem, is only a proper variable to use for
describing the ecosystem in the panmictic case.  One might attempt to
refine the concept of global frequency by including local frequencies.
However, the breakdown of moment-closure techniques implies that
defining fitness as a function of organism type together with average
local environment~\cite{smith2010} will, in many circumstances, not be
an adequate solution.

Consequently, we find that trying to assign a meaningful invasion
fitness value to an invasive variety of organism is too drastic a
simplification.  In turn, this implies that we cannot assign a fitness
value to a phenotypic or genetic characteristic such as infectiousness
or transmissibility.  To understand this point, we rephrase the
spatial host--consumer model in terms of alleles.  In an invasion
scenario, an individual consumer can have one of two possible alleles
of the ``transmissibility gene'', one coding for the native value
of~$\tau$ (\emph{e.g.,} $\tau = 0.33$) and the other for the invasive
value (\emph{e.g.,} $\tau = 0.45$).  A mean-field treatment would then
involve specifying the fraction of the population which carries the
native allele versus the fraction which carries the invasive variant.
We have seen, however, that the predictions based on such a heavily
coarse-grained caricature of the original model deviate from its
actual behavior.  In short, the evolutionary dynamics cannot be
characterized using the allele frequencies at a particular time.

If we can no longer summarize the genetic character of a population by
an allele frequency---or a set of allele frequencies for well-defined
local subpopulations---then computing the fitness of a genotype from
its generation-to-generation change in frequency is a fruitless task.
In a world which exhibits nonequilibrium spatial pattern formation,
allele frequencies are the wrong attribute for understanding the dynamics
of natural selection.  Formally, the conventional assumption that the
allele frequencies are a sufficient set of variables to describe
evolutionary dynamics is incorrect.  The spatial structure itself is a
necessary part of the system description at a particular time in order
to determine the subsequent generation outcomes, even in an average
sense.

The timescale-dependence issues which arise in spatial host--consumer
ecosystems exist in a wider context.  Multiple examples indicate that
initial success and eventual fixation are only two extremes of a
continuum which must be understood in its entirety to grasp the
stability of a system.  In the study of genetic drift, it has been
found that neutral mutations can fixate and beneficial mutations fail
to fixate due to stochasticity~\cite{hartl2007}.  Likewise, in the
study of clonal interference~\cite{fogle2008}, one beneficial mutation
can out-compete another and prevent its fixation.

Furthermore, classical genetics makes much use of the Price Equation
for studying the change in a population's genetic composition over
time~\cite{bijma2008, damore2011}, and it is well known that analytic
models built using the Price Equation lack ``dynamic
sufficiency''. That is, the equation requires more information about
the current generation than it produces about the
next~\cite{grafen1985, page2002, damore2011, simon2012, allen2012},
and so predictions for many-generation phenomena must be made
carefully, if they can be made at all.  Modeling approaches which are
fundamentally grounded in the Price Equation, such as
``neighbor-modulated'' fitness calculations~\cite{reeve2007,
  bijma2008, wild2009, lion2010, allen2013b} and their ``multilevel''
counterparts~\cite{reeve2007, bijma2008, wade2010, vandyken2011,
  sfd2012, nielsen2012}, are not likely to work well here, as the
analyses in question draw conclusions only from the short-timescale
regime.  In addition, those particular analyses which address
host--consumer-like dynamics either rely on moment
closures~\cite{lion2010} or they assume a fixed, complete connection
topology of local populations which are internally
well-mixed~\cite{wild2009, wade2010}.  These simplified population
structures are quite unlike the dynamical patch formation seen in the
host--consumer lattice model.  (Wild and Taylor~\cite{wild2004}
demonstrate an equivalence between stability criteria defined via
immediate gains, or ``reproductive fitness'', and criteria defined
using fixation probability; however, their proofs are explicitly
formulated for the case of a well-mixed population of constant size,
neither assumption being applicable here.  Whether fixation
probability is equivalent to any other criterion of evolutionary
success generally depends on mutation rates, even in
panmixia~\cite{allen2012}.)

In the adaptive dynamics literature, models have been studied in which
``the resident strikes back''~\cite{mylius2001, geritz2002,
  geritz2005}.  That is, an initially rare mutant variety $M$ can
invade a resident population of type $R$, but $M$ does not supplant
$R$ and become the new resident variety, even though a population full
of type $M$ is robust against incursions by type $R$.  This is often
considered a rare occurrence, requiring special conditions to
obtain~\cite{geritz2002, geritz2005}, though the theorems proved to
that effect apply to nonspatial models, and in adaptive dynamics, it
is standard to consider small differences between mutant and resident
trait values.  The spatial host--consumer ecosystem has the important
property that, if mutation is an ongoing process, the spatial extent
allows genetic diversity to grow.  We initialize the system with all
the consumers having the same trait value, but soon enough, different
local subpopulations have different trait values.  If the effects of
single mutations are small, then the different varieties arising have
roughly comparable survival probabilities, and so the distribution of
extant trait values can spread out.  However, the cumulative effect of
many mutations which happen to act in the same direction on a trait
such as transmissibility creates a variety which may engender its own
local Malthusian catastrophe.  So, the results of rare, big mutations
tell us about the spread of trait values we see in the case of
frequent, small mutations.

In our model, transmissibility and consumer death rate are
independently adjustable parameters. One can also build a model in
which one of these quantities is tied to the other, for example by
imposing a tradeoff between transmissibility and virulence of a
disease. Different functional forms of such a relationship are
appropriate for modeling different ecosystems: host/pathogen,
prey/predator, sexual/parthenogenetic and so forth. As long as spatial
pattern formation occurs and organism type impacts on the environment
of descendants via ecosystem engineering, the shortcomings of
mean-field theory are relevant, as are limitations of pair
approximations~\cite{messinger2012}.

\subsection{Pair Approximations and Stability}
\label{sec:discuss-pa}
Pair approximations have been used to test for the existence of an
{\em Evolutionary Stable Strategy} (ESS) in a system---that is, a
strategy which, when established, cannot be successfully replaced by
another~\cite{szabo2007}.  In addition to the limitations of pair
approximation for representing patch structure~\cite{vanbaalen2000},
as we saw in the previous section, the question of whether a mutant
strain can initially grow is distinct from the question of whether
that strain achieves fixation or goes extinct~\cite{doebeli1997,
  lessard2005, antal2005, strayer2006, paley2007, fogle2008,
  ponciano2009, heilmann2010, smaldino2013}.  The former is a question
about short-term behavior, and the latter concerns effects apparent at
longer timescales.  This distinction is often lost or obscured in
analytical treatments.  The reason is that one typically tests whether
a new type can invade by linearizing the corresponding differential
equations at a point where its density is negligible. However, this
only reveals the initial growth rate (see the fixed-point eigenvalue
analysis in~\cite{vanbaalen1998, vanbaalen2000, aguiar2004,
  lion2008}).

Our analysis implies that pair approximations are inadequate for
analysis of systems with spatial inhomogeneity.  Even including
including triple and other higher-order corrections does not suffice,
as this series approximation is poorly behaved at phase
transitions~\cite{buice2009}.  Such higher order terms continue to
reflect only the local structure of the system and not the existence
of well separated areas that diverge in their genetic composition.
Nonequilibrium pattern formation will necessarily also be poorly
described, at least until the order of expansion reaches the
characteristic number of elements in a patch, or an area that
encompasses any relevant heterogeneity.  Given the algebraic intricacy
of higher-order corrections to pair approximations~\cite{aguiar2004,
  lion2009, raghib2010}, it is useful to know in advance whether such
elaborations have a chance at success.  As approximation techniques
based on successively refining mean-field treatments are blind to
important phenomena, then we need to build our analytical work on a
different conceptual foundation.

\subsection{Percolation}
\label{sec:discuss-percolation}
The mathematical connection between pathogen--host and percolation
problems can provide insight into the difficulties in analytical
treatment of the biological problem. Spatial heterogeneity gives rise
to failure of traditional analytic treatments of percolation and a
need for new methodologies. Since the pathogen problem maps onto the
percolation problem under some circumstances, the same analytic
problems must arise in the biological context.  While the presence of
a nonequilibrium transition point indicates that traditional analysis
techniques fail, it raises the possibility that new tools from the
theory of phase transitions~\cite{hinrichsen2000, janssen2005,
  henkel2008, buice2009} will become applicable.  For example, in
Section~\ref{sec:percolation}, we saw that percolation theory enables
us to make quantitatively accurate predictions of population growth
and of the critical parameter values which divide one ecological
regime from another.  Indeed, specific important problems in public
health, such as the growth in number of individuals infected in a
pandemic can be considered directly within the context of
percolation. Simulations of propagation on approximate of real world
networks may help provide accurate predictions, but the general
properties of disease propagation can be understood analytically.

\subsection{Adaptive Networks}
\label{sec:adaptive-networks}
Our results also have significance in the context of adaptive-network
research.  This field studies systems in which a network's wiring
pattern and the states of its nodes change in interrelated ways.
Prior modeling efforts have considered epidemics on adaptive networks,
where the spread of the disease {\em through} the network changes the
connections {\em of} the network~\cite{gross2006, do2009, shaw2009,
  shaw2010, kamp2010, vansegbroeck2010, wu2010, fehl2011}.  In such
models, if a susceptible node has an infected neighbor, it can break
that connection by rewiring to another susceptible node.  A key point
in the analysis is that the new neighbor is chosen at random from the
eligible population.  This choice of rewiring scheme is exactly what
makes a pair approximation work for that epidemic model, because it
eliminates higher-order correlations in the system~\cite{do2009}.  In
our system, by contrast, hosts can form new connections by reproducing
into empty sites, but these contacts can only connect geographically
proximate individuals.

The difference we have seen between lattice behavior on one hand and
RRG or swapping-enabled behavior on the other emphasizes the need to
study the effect of spatial proximity on link rewiring.  While the
structure-erasing nature of unconstrained rewiring among susceptible
hosts has been acknowledged~\cite{kamp2010, vansegbroeck2010}, new
rewiring rules which reflect spatial and community structure have yet
to be systematically investigated.  The reason that they have not is
naturally related to the need for different analytic approaches.
``Myopic'' rewiring rules, such as restricting the set of eligible new
partners to the neighbors of a node's current partners, have on
occasion been considered, but in contexts other than epidemiology,
like evolutionary game theory~\cite{traulsen2009, holme2009}, making
the endeavour of exploring such rules in epidemic models all the more
worth pursuing.

\subsection{Conclusions}
Fisher \cite{fisher1930} introduced modern genetic theory in large
part motivated by the need to describe the existence of biodiversity.
However, the expressions he described which apply in panmictic
populations and mean-field treatments lead to a population genetics
that rapidly converges to homogeneous populations.  Spatial extents
and their violation of the mean-field approximations are a key to
biodiversity in nature.  Their proper theoretical treatment will be a
large step forward for evolutionary biology.

Most laboratory experiments, guided by traditional evolutionary
thinking, have used well-mixed populations.  The results obtained are
consistent with theoretical analysis precisely because the conditions
are consistent with those assumptions.  Such experiments do not
provide insight into the role of spatial extent and the implications
for real-world biological populations.  A growing number of
experiments today are going beyond such conditions and, as is to be
expected, are obtaining quite different results~\cite{kerr2006,
  xavier2007, davis2008, ponciano2009, heilmann2010, goenci2010,
  salkeld2010, reigada2012, wodarz2012, allen2013}.

Mean-field models are often helpful as a first step towards
understanding the behavior of systems, but we cannot trust them to
provide a complete story, and we should not let mean-field thinking
furnish all the concepts we use to reason about evolutionary dynamics.
Our analysis of transplanting organisms can be considered parallel to
real world concerns and manifest effects of invasive species
introduced by human activity and the impact of shipping and air
transportation on pathogen evolution~\cite{strayer2006, rauch2006}.
These are among the well-established examples of situations in which
spatial extent influences evolutionary dynamics~\cite{tgoei2000,
  ponciano2009, aguiar2009, goenci2010, aguiar2011, martins2013,
  vandyken2013, baptestini2013}.  Identifying specific implications of
the issues explored in this paper for particular biological
systems~\cite{kerr2006, xavier2007, davis2008, ponciano2009,
  heilmann2010, goenci2010, salkeld2010, reigada2012, wodarz2012,
  allen2013} requires field and laboratory work, as well as
theoretical insight to guide the questions that are being asked.

\section{Acknowledgements}

We thank M.\ de Aguiar, B.\ Allen, K.\ Z.\ Bertrand, E.\ Downes,
A.\ S.\ Gard-Murray and D.\ Harmon for providing helpful comments on
the manuscript.

\bibliography{report-redgreen.bib}

\end{document}